\def\etal{{\frenchspacing\it et al.}}

\def\beq#1{\begin{equation}\label{#1}}
\def\eeq{\end{equation}}
\def\beqa#1{\begin{eqnarray}\label{#1}}
\def\eeqa{\end{eqnarray}}

\def\ztwodf{z_{\rm 2df}}
\def\etal{{\frenchspacing\it et al.}}

\documentclass[12pt,preprint]{aastex}

\usepackage{emulateapj5}

\usepackage{graphicx}
\usepackage{epsfig}

\newcommand{\be}{\begin{equation}}
\newcommand{\ee}{\end{equation}}
\newcommand{\ba}{\begin{eqnarray}}
\newcommand{\ea}{\end{eqnarray}}

\shorttitle{Robust Dark Energy Constraints}
\shortauthors{Wang \& Mukherjee}

\begin{document}

\title{Robust Dark Energy Constraints from Supernovae, 
Galaxy Clustering,
and Three-Year Wilkinson Microwave Anisotropy Probe Observations}
\author{Yun~Wang$^{1}$, \& Pia~Mukherjee$^{2}$}
\altaffiltext{1}{Department of Physics \& Astronomy, Univ. of Oklahoma,
                 440 W Brooks St., Norman, OK 73019;
                 email: wang@nhn.ou.edu}
\altaffiltext{2}{Department of Physics \& Astronomy, Univ. of Sussex, 
Falmer, Brighton, BN1 9QH; email: p.mukherjee@sussex.ac.uk}

\begin{abstract}

Type Ia supernova (SN Ia), galaxy clustering, and 
cosmic microwave background anisotropy (CMB) data provide complementary 
constraints on the nature of the dark energy in the universe. 
We find that the three-year Wilkinson Microwave Anisotropy Probe (WMAP) 
observations give a CMB shift parameter of $R \equiv
\left(\Omega_m H_0^2\right)^{1/2} \int_0^{z_{CMB}} dz'/H(z')=
1.70 \pm 0.03$.
Using this new measured value of the CMB shift parameter,
together with the baryon acoustic oscillation (BAO) measurement 
from the Sloan Digital Sky Survey (SDSS),
and SN Ia data from the HST/GOODS program and the first year Supernova 
Legacy Survey, we derive model-independent constraints on the dark energy 
density $\rho_X(z)$ and the cosmic expansion rate $H(z)$.
We also derive constraints on the dark energy 
equation of state $w_X(z)=w_0+w'z$ (with cutoff at $z=2$)
and $w_X(a)=w_0+(1-a)w_a$.

We find that current data provide slightly tighter constraints 
on $\rho_X(z)$ and $H(z)$ as free functions in redshift, and roughly a 
factor of two improvement in constraining $w_X(z)$.
A cosmological constant remains consistent with data,
however, uncertainties remain large for model-independent constraints
of dark energy. Significant increase in the number of observed SNe Ia between
redshifts of 1 and 2, complemented by improved BAO and weak lensing 
cosmography measurements (as expected from the JEDI mission concept 
for the Joint Dark Energy Mission),
will be required to dramatically 
tighten model-independent dark energy constraints. 

\end{abstract}


\keywords{Cosmology}

\section{Introduction}

The observed accelerated expansion of the universe \citep{Riess98,Perl99}
can be explained by an unknown energy component in
the universe \citep{Freese87,Linde87,Peebles88,Wett88,Frieman95,Caldwell98},
or a modification of general relativity 
\citep{SH98,Parker99,DGP00,Mersini01,Freese02}.
\cite{Pad} and \cite{Peebles03} contain reviews of many models.
Some recent examples of models are presented in 
\cite{Carroll04,OW04,Cardone05,Kolb05,MB05,McInnes05,Cai06}.
For convenience, we refer to the cause for the cosmic acceleration
as ``dark energy''.

Data of Type Ia supernovae (SNe Ia), cosmic large scale structure (LSS),
and the cosmic microwave anisotropy (CMB) are complementary
in precision cosmology \citep{Bahcall99,Eisen99,Wang99}.
An important development in this complementarity is to
use the baryonic acoustic oscillations (BAO) in the galaxy power
spectrum as a cosmological standard ruler to probe dark energy \citep{BG03,SE03}.

In placing robust constraints on dark energy, it is important to
(1) derive model-independent dark energy constraints 
\citep{WangGarnavich01,Tegmark02,Daly03},
and (2) use data derived {\it without} assuming dark energy to be 
a cosmological constant \citep{WangTegmark04}. 

In this paper, we derive the CMB shift parameter from the
three-year Wilkinson Microwave Anisotropy Probe (WMAP) 
observations \citep{Spergel06,Bennett03}, and show that
its measured value is mostly independent of assumptions made
about dark energy.
Using this new measured value of the CMB shift parameter,
together with the LSS data from the BAO measurement from the 
Sloan Digital Sky Survey (SDSS),
and SN Ia data from the HST/GOODS program and the first year Supernova 
Legacy Survey (SNLS), we derive model-independent constraints on the dark
energy density $\rho_X(z)$ and the cosmic expansion rate $H(z)$.
For reference and comparison,
we also derive constraints on the linear dark energy 
equation of state $w_X(z)=w_0+w'z$ (with cutoff at $z=2$)
and $w_X(a)=w_0+(1-a)w_a$ \citep{Cheva01}.

Sec.2 describes the method and data used in our calculations.
We present results in Sec.3 and summarize in Sec.4.

\section{The method and data used}

\subsection{The method}

We run a Monte Carlo Markov Chain (MCMC) based on the MCMC engine 
of \cite{Lewis02} to obtain ${\cal O}$($10^6$) samples for each set of 
results presented in this paper. The chains are subsequently 
appropriately thinned.

We derive constraints on the dark energy density $\rho_X(z)$
as a free function, with its value at redshifts
$z_i$, $\rho_X(z_i)$, treated as independent parameters estimated from data.
For $z>z_{\rm cut}$, we assume $\rho_X(z)$ to be a powerlaw smoothly
matched on to $\rho_X(z)$ at $z=z_{\rm cut}$ \citep{WangTegmark04}:
\be
\rho_X(z)=\rho_X(z_{\rm cut}) \left( \frac{1+z}{1+z_{\rm cut}}\right)^{\alpha}.
\ee
The number of observed SNe Ia is either very few or none beyond $z_{\rm cut}$.
For the \cite{Riess04} sample, $z_{\rm cut}=1.4$. 
For the \cite{Astier05} sample, $z_{\rm cut}=1.01$. 
We use cubic spline interpolation to obtain values of $\rho_X(z)$ at other
values of $z$ \citep{WangTegmark04}. 

The $H(z)$ values corresponding to the estimated $\rho_X(z_i)$
are estimated directly from the MCMC chain to fully incorporate
the correlation between the estimated parameters, and compared
with the uncorrelated estimates of $H(z)$ from SN Ia data only 
\citep{WangTegmark05}.

For reference and comparison with the work by others, we
also derive constraints on dark energy models with 
a constant dark energy equation of state $w$ ($w_X(z)=w$),
and a linear equation of state parametrized by 
(1) $w_X(z)=w_0+w'z$ at $z\leq 2$, and
$w_X(z)=w_0+2w'$ at $z>2$;
(2) $w_X(a)=w_0+(1-a)w_a$.

\subsection{SN Ia data}

Calibrated SN Ia data \citep{Phillips93,Riess95} give luminosity distances 
$d_L(z_i)$ to the redshifts of the SNe Ia $z_i$. For a flat universe
\be
\label{eq:d_L}
d_L(z)=cH_0^{-1} (1+z) \int_0^z \frac{{\rm d}z'}{E(z')},
\ee
where
\be
E(z)\equiv \left[\Omega_m (1+z)^3 + (1-\Omega_m) \rho_X(z)/\rho_X(0)\right]^{1/2},
\ee
with $\rho_X(z)$ denoting the dark energy density.

We use SN Ia data from the HST/GOODS program \citep{Riess04} and the first 
year SNLS \citep{Astier05}, together with nearby SN Ia data. The comparison of 
results from these two data sets provides a consistency check. 

We do not
combine these two SN Ia data sets, as they have systematic differences
in data processing. Fig.1 shows the difference in estimated distance
moduli for 37 SNe Ia included in both the \cite{Riess04} ``gold'' sample
and the \cite{Astier05} sample; there is clearly significant scatter due to difference
in analysis techniques \citep{Wang00b}.
As a result, the two data sets have noticably different zero point  
calibrations. A given zero point calibration affects the measuremnt of $H_0$, but
has no impact on dark energy constraints (see Eq.[\ref{eq:d_L}]).
Combining the data sets will lead to artificial systematic errors 
resulting from the difference in zero point calibrations, which are difficult
to quantify; this outweighs the gain in accuracy at present since 
there is a large overlap between the two data sets.
It is important to use data analyzed using the same technique (which corresponds to
the same zero point calibration); although it would be useful to use the same
data analyzed using one or more other techniques (but only one technique should 
be used at a time) for cross check.

We use the \cite{Riess04} ``gold'' sample flux-averaged with $\Delta z=0.05$.
This sample includes 9 SNe Ia at $z>1$, and appears to have systematic effects 
that would bias the distance estimates somewhat without flux-averaging
\citep{WangTegmark04,Wang05}.
Flux-averaging \citep{Wang00b,WangMukherjee04}
removes the bias due to weak lensing magnification of SNe Ia 
\citep{Kantow95,frieman97,Wamb97,Holz98,ms99,Wang99a,Barber00,
Vale03}\footnote{Weak lensing magnification of SNe Ia
can also be used as a cosmological probe, see 
\cite{Dodelson05,Cooray06,Munshi06}.},
or other systematic effects that mimics weak lensing qualitatively in affecting
the observed SN Ia brightness.

We have added a conservative estimate of 0.15 mag in intrinsic dispersion 
of SN Ia peak brightness in quadrature to the distance moduli published 
by \cite{Astier05}, instead of using 
the smaller intrinsic dispersion derived by \cite{Astier05} by requiring
a reduced $\chi^2=1$ in their model fitting. 
This is because the intrinsic dispersion in SN Ia peak brightness 
should be derived from the distribution of nearby SNe Ia, or SNe Ia
from the same small redshift interval if the distribution in
SN Ia peak brightness evolves with cosmic time. 
This distribution is not well known at present, but will become better
known as more SNe Ia are observed by the nearby SN Ia factory \citep{Aldering02}
and the SNLS \citep{Astier05}.
By using the larger intrinsic dispersion,
we allow some margin for the uncertainties in SN Ia peak brightness distribution.

\subsection{LSS data}

For LSS data, we use the measurement of the BAO peak in
the distribution of SDSS luminous red galaxies (LRG's).
We do not use the linear growth rate measured by the 2dF survey,
as there are some inconsistencies in currently published
2dF results \citep{Verde02,Hawkins03}.\footnote{Combining the results
from \cite{Verde02} and \cite{Hawkins03}, we would obtain
the linear growth rate $f(\ztwodf)=0.51\pm 0.11$. However,
\cite{Hawkins03} points out that the \cite{Verde02} results
strongly depend on the assumed pairwise peculiar velocity dispersion
of 385$\,{\rm km/s}$, while \cite{Hawkins03} finds the 
pairwise peculiar velocity dispersion to be 500$\,{\rm km/s}$.
A more self-consistent linear growth rate from 2dF data has to await
a new bispectrum analysis of the 2dF data.}

The SDSS BAO measurement \citep{Eisen05} gives 
$A=0.469\,(n_S/0.98)^{-0.35}\pm 0.017$
(independent of a dark energy model) at $z_{BAO}=0.35$, where $A$ is defined as
\be
\label{eq:A}
A = \left[ r^2(z_{BAO})\, \frac{cz_{BAO}}{H(z_{BAO})} \right]^{1/3} \, 
\frac{\left(\Omega_m H_0^2\right)^{1/2}} {cz_{BAO} },
\ee
where $r(z)$ is the comoving distance, and $H(z)$ is the Hubble parameter.
Note that $H(z)=H_0 E(z)$.
We take the scalar spectral index $n_S=0.95$ as measured by WMAP3
\citep{Spergel06}.
Note that this constraint from \cite{Eisen05} 
is not just a simple measurement of the BAO feature; it also relies
on the constraints on $\Omega_m h^2$ from 
measuring the power spectrum turnover scale (related to matter-radiation equality). 
The latter makes their BAO constraint less robust than it would be otherwise.
A new analysis of the SDSS data to derive truly robust BAO constraints would
be very useful for placing dark energy constraints \citep{Dick06}.

Also note that the \cite{Eisen05} constraint on $A$ depends on the 
scalar spectral index $n_S$. Since the error on $n_S$ from WMAP data does not 
increase the effective error on $A$, and the correlation between $n_S$ and
the CMB shift parameter $R$ is weak, we have ignored the very weak correlation
between $A$ and $R$ in our likelihood analysis.
We have derived $R$ from WMAP data marginalized over all relevant parameters.

\subsection{CMB data}

The CMB shift parameter $R$ is perhaps the
least model-dependent parameter that can be extracted from CMB data,
since it is independent of $H_0$. The shift parameter $R$ is given by \citep{Bond97}
\be
\label{eq:R}
R\equiv {\Omega_m}^{1/2}\int_0^{z_{CMB}} {\rm d}z'/E(z'),
\ee
where $z_{CMB}$ is the redshift of recombination; 
thus $R= \left({\Omega_m H_0^2}\right)^{1/2} r(z_{CMB})/c$
(for a flat universe),
which is well determined since both $\Omega_m h^2$ and 
$r(z_{CMB})$ are accurately determined by CMB data.
Similar reasoning applies to an open and a closed universe
as well. 

The ratio of the sound horizon at recombination to $r(z_{CMB})$,
$\theta_s$, should be more or less equivalent to $R$. We call 
$R$ the ``CMB shift parameter'' following previously published literature 
over the last 9 years (see for example, \cite{Bond97,Odman03}). One
advantage of $R$ is that it only involves a simple integral over $1/E(z)$, while
the sound horizon at last scattering is more complicated to calculate accurately
and depends on more parameters. 

We compute $R$ using the MCMC chains from the analysis of the 
three year WMAP data provided by the WMAP team \citep{Spergel06}.
The resultant probability distribution of $R$ for three
different classes of models are shown in Fig.2. Clearly, the measured $R$ 
from WMAP 3 year data has only a very weak model dependence\footnote{For a
 Bayesian analysis of the number of parameters required by current 
cosmological data, in other words of what comprises an adequate model,
 see \cite{Mukherjee06}}.

We use a Gaussian distribution in $R$ with $\langle R\rangle=1.70$ and
$\sigma_R=0.03$ (thin solid line in Fig.2) in deriving our results
presented in Sec.3. This fits the WMAP 3 year data
well, and allows some margin in error for the very weak model dependence
of $R$. 

\section{Results}

In deriving all the results presented in this section, we have assumed
a flat universe, and marginalized over $\Omega_m$ and $h$.

Fig.3 shows $\rho_X(z)$ measured using SN Ia data \citep{Riess04,Astier05},
combined with the WMAP 3 year data, and the SDSS BAO data.
Note that beyond $z_{\rm cut}$=1.4 (upper panel)
and 1.01 (lower panel), $\rho_X(z)$ is parametrized by a power law
$(1+z)^\alpha$, the index of which is marginalized over.

Fig.4 shows the uncorrelated $H(z)$ estimated using only SN Ia data
\citep{WangTegmark05},
and the $H(z)$ corresponding to the $\rho_X(z_i)$ shown in Fig.3.

Tables 1 and 2 give the 68\% confidence intervals of 
$\rho_X(z_i)$, $\alpha$, and $H(z_i)$ for the 
WMAP 3 year and the SDSS BAO data, 
combined with SN Ia data from \cite{Riess04}
and \cite{Astier05} respectively.

\begin{table*}[htb]
\caption{The mean and the 68\% confidence intervals of 
$\rho_X(z_i)$, $\alpha$, and $H(z_i)$ for
the CMB and LSS data 
combined with SN Ia data from \cite{Riess04}.}
\begin{center}
\begin{tabular}{c|c}
\hline\hline

Parameter & Riess04+WMAP3+SDSS \\
\hline
$\rho_X(0.467)$ & 1.159  (0.953, 1.361)\\
$\rho_X(0.933)$ & 1.357  (0.602, 2.095)\\
$\rho_X(1.400)$ & 2.751  (0.408, 5.053)\\
$\alpha$ & -0.037 (-1.927, 1.818)\\
\hline
$H(0.467)$ & 1.321 (1.257, 1.384)\\
$H(0.933)$ & 1.758 (1.617, 1.896)\\
$H(1.400)$ & 2.419 (2.045, 2.797)
 \\
\hline
 \hline		
\end{tabular}
\end{center}
\end{table*}

\begin{table*}[htb]
\caption{The mean and the 68\% confidence intervals of 
$\rho_X(z_i)$, $\alpha$, and $H(z_i)$ for
the CMB and LSS data 
combined with SN Ia data from \cite{Astier05}.}
\begin{center}
\begin{tabular}{c|c}
\hline\hline

Parameter & Astier05+WMAP3+SDSS \\
\hline
$\rho_X(0.505)$ & 1.013  (0.893, 1.131)\\
$\rho_X(1.010)$ & 1.579  (0.833, 2.327)\\
$\alpha$ & 0.030  (-1.786, 1.770)\\
\hline
$H(0.505)$ & 1.290  (1.252, 1.327)\\
$H(1.010)$ & 1.828   (1.678, 1.978)\\

\hline

\hline		
\end{tabular}
\end{center}
\end{table*}

Tables 3 and 4 give the covariance matrices for ($\rho_X(z_i)$, $\alpha$),
and ($H(z_i)$, $\alpha$) for the WMAP3 and SDSS BAO data combined with 
SN Ia data from \cite{Riess04}
and \cite{Astier05} respectively.

\begin{table*}[htb]
\caption{The covariance matrices for 
$\rho_X(z_i)$ and $H(z_i)$ from
the WMAP 3 year and the SDSS BAO data
combined with SN Ia data from \cite{Riess04}.}
\begin{center}
\begin{tabular}{|c|llll|}
\hline
&$\rho_X(0.467)$ & $\rho_X(0.933)$ &$\rho_X(1.400)$ & $\alpha$ \\
\hline
$\rho_X(0.467)$ & 0.426E-01 & 0.226E-01 & -0.137E+00 & 0.153E-01\\
$\rho_X(0.933)$ &  0.226E-01 & 0.585E+00 & -0.261E+00 & -0.322E+00 \\
$\rho_X(1.400)$ & -0.137E+00 & -0.261E+00 & 0.717E+01 & -0.110E+01 \\
$\alpha$ &  0.153E-01 & -0.322E+00 & -0.110E+01  & 0.276E+01  \\

  \hline
  \hline
  &$H(0.467)$&$H(0.933)$ &$H(1.400)$ & $\alpha$ \\
  \hline
  $H(0.467)$ & 0.407E-02  & 0.185E-02 & 0.121E-02 & 0.851E-02  \\     
  $H(0.933)$&  0.185E-02 & 0.200E-01 & -0.130E-02 & -0.534E-01    \\   
  $H(1.400)$ & 0.121E-02 & -0.130E-02 & 0.154E+00 & -0.125E+00  \\   
  $\alpha$ & 0.851E-02 & -0.534E-01&  -0.125E+00 & 0.276E+01  \\
  
  \hline
\end{tabular}
\end{center}
\end{table*}

\begin{table*}[htb]
\caption{The covariance matrices for 
$\rho_X(z_i)$ and $H(z_i)$ from
the WMAP 3 year and the SDSS BAO data
combined with SN Ia data from \cite{Astier05}.}
\begin{center}
\begin{tabular}{|c|lll|}
\hline
&$\rho_X(0.505)$ & $\rho_X(1.010)$ & $\alpha$ \\ 
\hline
$\rho_X(0.505)$ & 0.148E-01 & 0.472E-01 & -0.750E-01  \\
$\rho_X(1.010)$ &  0.472E-01 & 0.559E+00 & -0.460E+00  \\
$\alpha$ & -0.750E-01 & -0.460E+00 & 0.254E+01\\
  
  \hline
  \hline
  &$H(0.505)$&$H(1.010)$ & $\alpha$ \\
  \hline
  $H(0.505)$ &  0.146E-02 & 0.308E-02 & -0.118E-01 \\    
  
  $H(1.010)$&   0.308E-02 & 0.223E-01 & -0.729E-01 \\  
    
  $\alpha$ & -0.118E-01 & -0.729E-01 & 0.254E+01  \\ 
  
  \hline
\end{tabular}
\end{center}
\end{table*}

Table 5 gives the constraints on a constant $w$ ($w_X(z)=$const.), 
($w_0,w')$ ($w_X(z)=w_0+w'z$ at $z\leq 2$, and
$w_X(z)=w_0+2w'$ at $z>2$), and ($w_0,w_a)$ ($w_X(a)=w_0+(1-a)w_a$).

\begin{table*}[htb]
\caption{The mean and 68\% and 95\% confidence level constraints on 
a constant $w$ and ($w_0,w')$, and the covariance
between $w_0$ and $w'$, and between $w_0$ and $w_a$.}
\begin{center}
\begin{tabular}{c||c|c}
\hline
& {\rm Riess04+WMAP3+SDSS} & {\rm Astier05+WMAP3+SDSS} \\
\hline
\hline
$w$ & -0.885$\, _{-0.111}^{+0.109} \, _{-0.227}^{+0.206}$ &

-0.999 $\,_{-0.083}^{+0.082}\, _{-0.168}^{+0.159}$ \\
\hline
\hline

$w_0$ & -0.794 $\, _{-0.244}^{+0.243} \, _{-0.431}^{+0.584}$&
 -0.989 $\, _{-0.162}^{+0.160} \, _{-0.291}^{+0.413}$\\
 \hline
$w'$ & -0.446 $\, _{-0.710}^{+0.711} \, _{-2.237}^{+0.948}$ & 

-0.177 $\, _{-0.574}^{+0.571} \, _{-1.984}^{+0.742}$\\
\hline

$\sigma^2(w_0,w')$ & -0.192 &  -0.112\\
\hline
\hline
$w_0$ & -0.813 $\, _{-0.296}^{+0.293} \, _{-0.508}^{+0.704}$&

-1.017 $\, _{-0.200}^{+0.199} \, _{-0.350}^{+0.503}$\\

\hline
$w_a$ & -0.510  $\, _{-1.259}^{+1.265} \, _{-3.620}^{+1.792}$ &

-0.039 $\, _{-1.052}^{+1.045} \, _{-3.173}^{+1.429}$\\
\hline
$\sigma^2(w_0,w_a)$ & -0.410 & -0.245\\
\hline
\hline

\end{tabular}
\end{center}
\end{table*}

Fig.5 shows the 68\% and 95\% joint confidence contours for $(w_0,w')$
and $(w_0, w_a)$.

\section{Discussion and Summary}

In order to place robust constraints on dark energy in a simple 
and transparent manner, we have derived the CMB shift parameter 
from the WMAP 3 year data ($R=1.70 \pm 0.03$, see Fig.2).
We constrain dark energy using this new measurement of the
CMB shift parameter, together with LSS data (the BAO measurement 
from the SDSS LRG's),
and SN Ia data (from the HST/GOODS program and the first year SNLS).

We have derived model-independent constraints on the dark energy 
density $\rho_X(z)$ and the cosmic expansion rate $H(z)$ (Figs.3-4
and Tables 1-4).
There are two reasons that one should use $\rho_X(z)$ and $H(z)$ 
instead of $w_X(z)$ to probe dark energy.
First, $\rho_X(z)$ and $H(z)$ are more directly related to observables
than $w_X(z)$ (see Eqs.[\ref{eq:d_L}]-[\ref{eq:R}],
and note that $\partial\ln\rho_X/\partial\ln a= -3(1+w_X)$)
\citep{WangGarnavich01,Tegmark02}.
This means that $\rho_X(z)$ is more tightly constrained by data
than $w_X(z)$ \citep{WangFreese06,Daly05,Huterer05,Ishak05,Dick06}. 
Secondly, $\rho_X(z)$ and $H(z)$ are more general phenomenological
representations of dark energy than $w_X(z)$ \citep{WangTegmark04}. 
One must integrate the equation,
$\partial\ln\rho_X/\partial\ln a= -3(1+w_X)$,
to obtain $\rho_X(z)$ before a 
comparison with data can be made, hence even arbitrary function 
$w_X(z)$ has the hidden assumption that $\rho_X(z)$ is non-negative 
(since $\rho_X(0)\geq 0$).
Since we don't know what dark energy is -- it may
not even be energy at all but a modification of general relativity -- 
$\rho_X(z)$ may well have been negative at some past epoch or become
negative in the future.
Measuring $\rho_X(z)$ and $H(z)$ (instead of $w_X(z)$) from data 
allows us to constrain this possibility.

For comparison with the work by others, 
we also derive constraints on the dark energy 
equation of state $w_X(z)=w_0+w'z$ (with cutoff at $z=2$)
and $w_X(a)=w_0+(1-a)w_a$ (Fig.5 and Table 5).

Because of the difference in SN Ia analysis techniques used in
deriving the \cite{Riess04} ``gold'' sample and \cite{Astier05} 
sample (see Fig.1), we have presented
the results for these two data sets separately. 
This is necessary in order to avoid introducing systematic errors
due to the different zero point calibrations of the two data sets
(which are difficult to quantify) \citep{Wang00b}.
We find that
the \cite{Riess04} and \cite{Astier05} data sets
give similar and consistent constraints on dark energy
(Figs.3-5), with the \cite{Riess04} sample being able to constrain
$\rho_X(z)$ and $H(z)$ at $z>1$ since it contains 9 SNe Ia at $z>1$.
This is reassuring, and seems to indicate that
the \cite{Riess04} sample (after flux-averaging) is consistent
with the \cite{Astier05} sample. It will be useful to derive
the distances to {\it all} the SNe Ia using the {\it same}
technique to reduce systematic uncertainties \citep{Wang00b}
and maximize the number of SNe Ia that can be used in the same dark energy
analysis.

We find that compared to previous results in \cite{WangTegmark04},
the current data provide a slightly tighter constraint on $\rho_X(z)$
and $H(z)$ as free functions of cosmic time, and roughly a 
factor of two improvement in constraining $w_X(z)$.
Note that the \cite{Astier05} data set (together with WMAP 3 year
and SDSS BAO data) provides a tighter constraint on 
$w_0 \equiv w_X(z=0)$ because it contains more SNe Ia at lower redshifts.
Because of the strong correlation between $w'$ (or $w_a$) and $w_0$,
this has led to tighter constraints on $w'$ and $w_a$ as well.

A cosmological constant remains consistent with data, although
more exotic possibilities are still allowed, consistent with previous results 
(see for example, 
\cite{WangTegmark04,WangTegmark05,Alam05,Daly05,Jassal05a,Jassal05b,Dick06,Ichi06,Jassal06,Nesseris06,Schimd06,Wilson06}).
In particular, \cite{Zhao06} uses WMAP 3 year, SN Ia data from 
\cite{Riess04}, together with SDSS 3D power spectra and Lyman-$\alpha$ 
forest information data, and found constraints on $w_X(z)=w_0+w_1z/(1+z)$
that are qualitatively consistent with our results, with significant differences 
that can be explained by the differences in the combination of data used,  
and perhaps some data analysis details. 

We find that uncertainties remain large for model-independent constraints
of dark energy (see Figs.3-4). Significant increase in the number 
of observed SNe Ia between redshifts of 1 and 2, complemented by BAO and 
weak lensing cosmography measurements (such as expected from the JEDI mission
concept of the Joint Dark Energy Mission), should dramatically
tighten dark energy constraints and
shed light on the nature of dark energy \citep{Wang00a,JEDI}.\footnote{We 
have assumed a flat universe for all our results.
When data of significantly larger quantity and higher quality 
become available from JDEM and the next generation ground-based surveys,
it will be possible to constrain the curvature of the universe as well.}

\bigskip

{\bf Acknowledgements}
It is a pleasure for us to thank Max Tegmark for helpful discussions,
and the referee for useful suggestions.
YW is supported in part by NSF CAREER grants AST-0094335. PM is 
supported by PPARC (UK).

\clearpage
\setcounter{figure}{0}

\begin{figure}
\centering
\includegraphics[width=11cm]{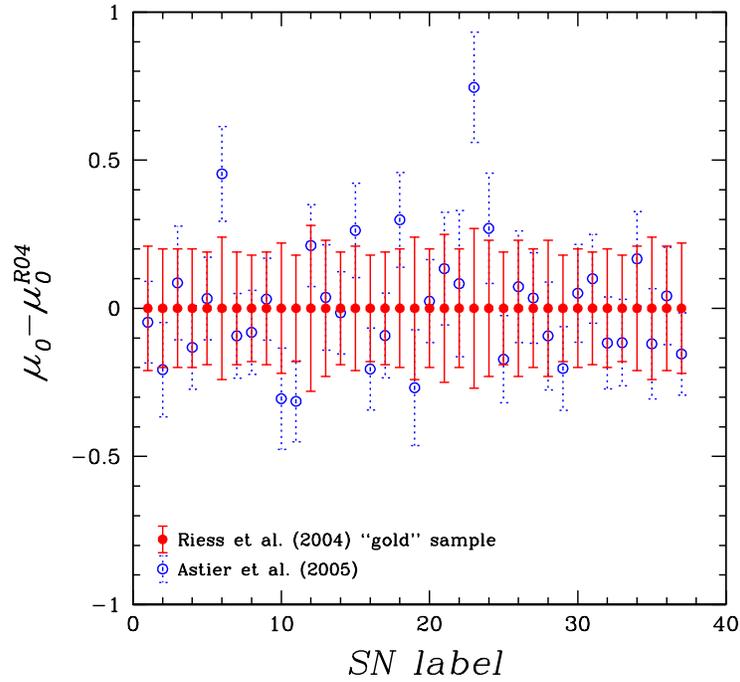}
\caption{The difference in estimated distance
moduli for 37 SNe Ia included by both the \cite{Riess04} ``gold'' sample
and \cite{Astier05}; there is clearly significant scatter due to difference
in analysis techniques \citep{Wang00b}.}
\label{fig1}
\end{figure}

\begin{figure}
\centering
\includegraphics[width=11cm]{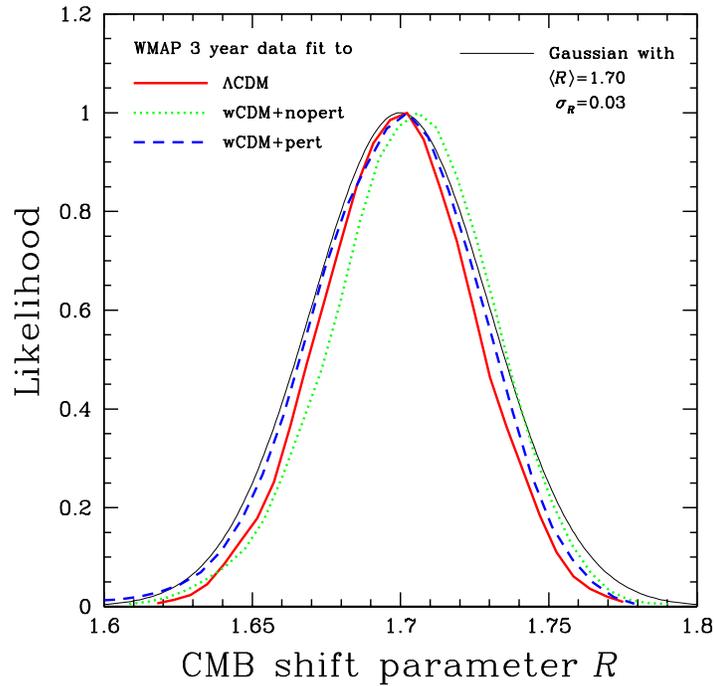}
\caption{The probability distribution of $R$ for three
different classes of models.}
\label{fig2}
\end{figure}

\begin{figure}
\centering
\includegraphics[width=16cm]{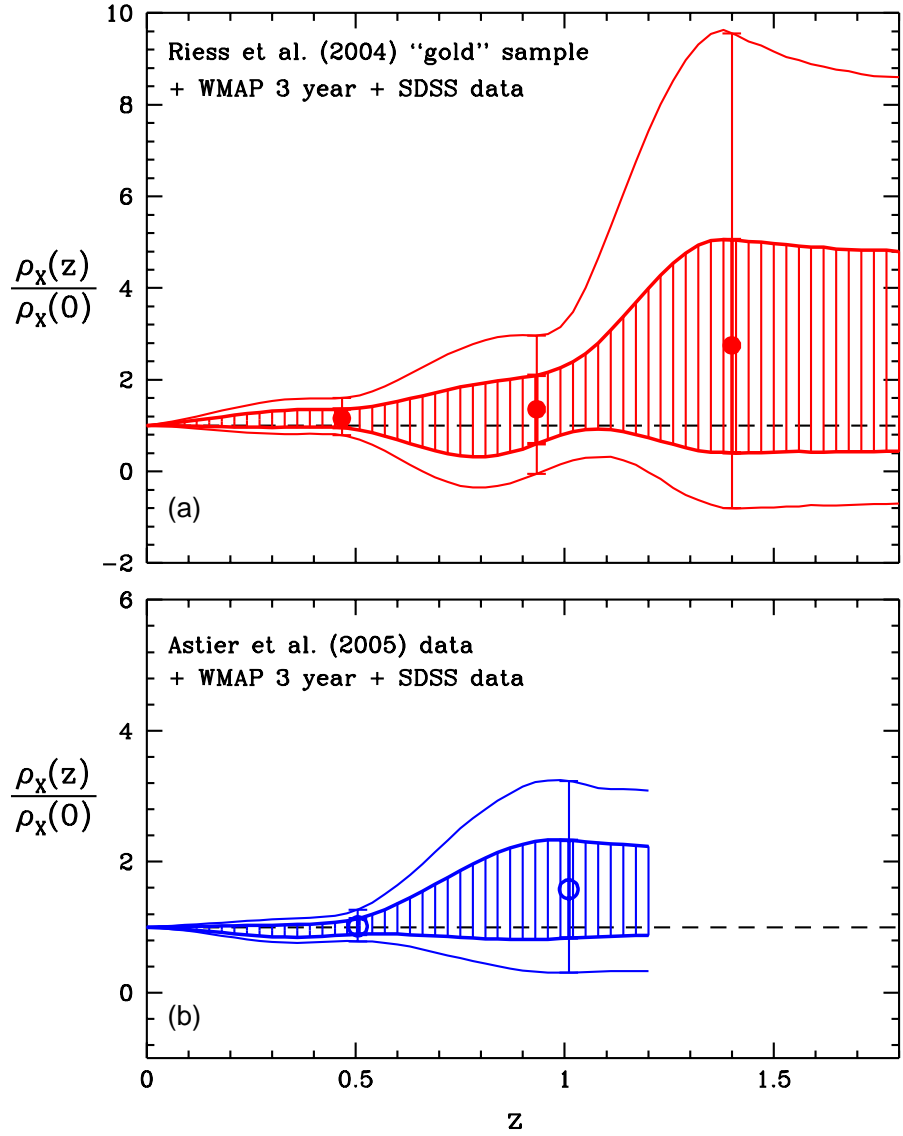}
\caption{Dark energy density $\rho_X(z)$ measured using SN Ia data \citep{Riess04,Astier05}, 
combined with the WMAP 3 year data, and the SDSS BAO data.
The 68\% (shaded) and 95\% confidence contours are shown.
Note that beyond $z_{\rm cut}$=1.4 (upper panel) and 1.01 (lower panel), 
$\rho_X(z)$ is parametrized by a power law $(1+z)^\alpha$.
}
\label{fig3}
\end{figure}

\begin{figure}
\centering
\includegraphics[width=14cm]{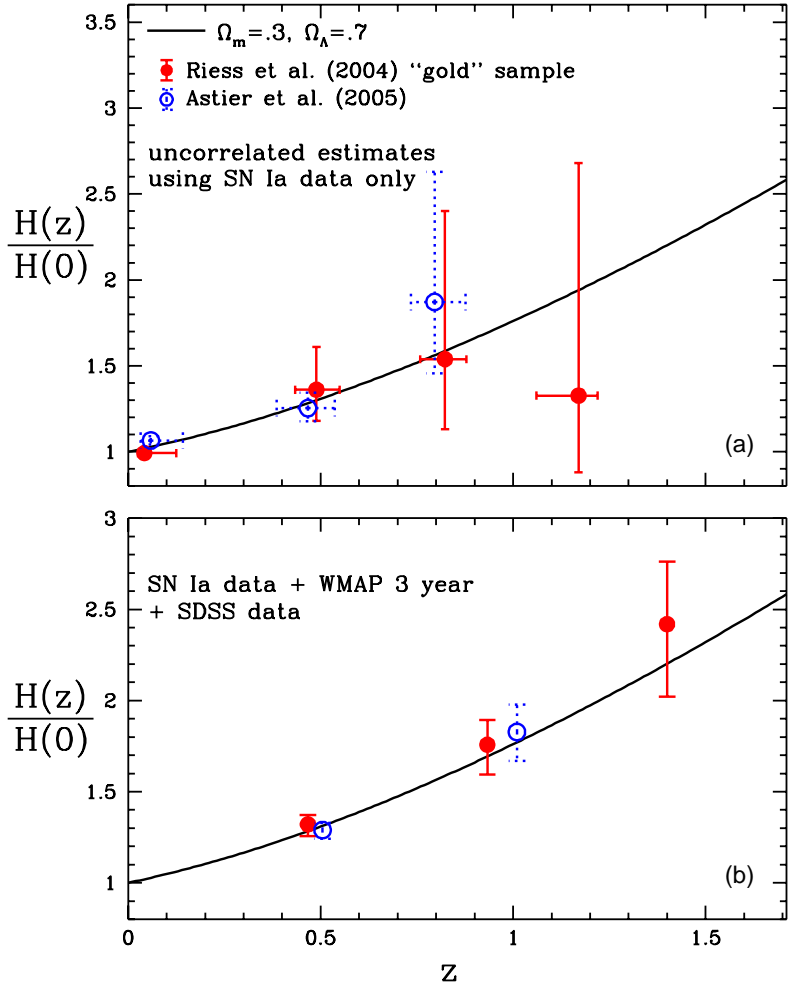}
\caption{Upper panel: the uncorrelated $H(z)$ estimated using only SN Ia data.
Lower panel: the $H(z)$ corresponding to the $\rho_X(z_i)$ shown in Fig.3.
The error bars indicate the 68\% confidence limits.
}
\label{fig4}
\end{figure}

\begin{figure}
\centering
\includegraphics[height=7cm]{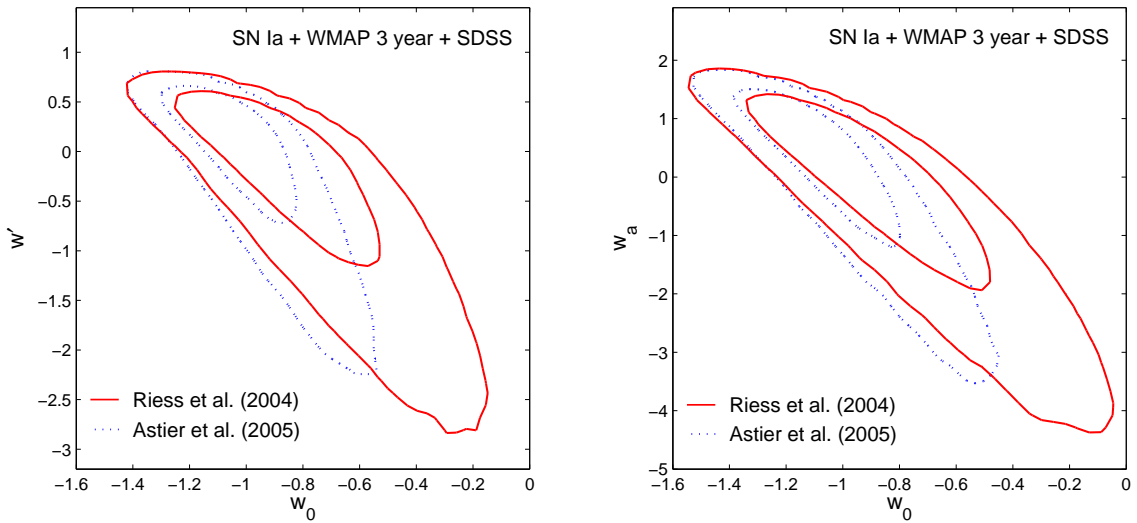}
\caption{The 68\% and 95\% joint confidence contours for $(w_0,w')$ and $(w_0,w_a)$.}
\label{fig5}
\end{figure}

\end{document}